\documentclass[12pt]{iopart}

\usepackage[colorinlistoftodos,prependcaption,textsize=tiny]{todonotes}
\usepackage{iopams}

\usepackage{mathrsfs}

\usepackage{diagbox}
\usepackage{booktabs}
\usepackage{braket}
\usepackage{xcolor}

\newcommand{\Het}[0]{He$_2^*$ }
\newcommand{\Hetwo}{He$_2$ }
\newcommand{\Hep}{He$_2^+$ }
\newcommand{\wn}{cm$^{-1}$}

\newcommand{\bb}{{\it b}~}
\newcommand{\cc}{{\it c}~}

\newcommand{\ee}{{\it e}~}

\newcommand{\eqref}[1]{(\ref{#1})}

\newcommand{\Wthreej}[6]{\left( \begin{array}{ccc} #1 & #2 & #3 \\ #4 & #5 & #6 \end{array} \right)}
\newcommand{\Wsixj}[6]{\left\{ \begin{array}{ccc} #1 & #2 & #3 \\ #4 & #5 & #6 \end{array} \right\}}

\begin{document}

\title{Laser cooling Rydberg molecules - a detailed study of the helium dimer}

\author{Lucía Verdegay$^1$, Bingcheng Zeng$^1$, Daniel Y. Knapp$^1$, Jack C. Roth$^2$ and
Maximilian Beyer$^1$}
\address{$^1$ Department of Physics and Astronomy,
 Vrije Universiteit Amsterdam, The Netherlands} 
\address{$^2$ Department of Physics, University of California, Berkeley, CA 94720, USA}

\ead{m.beyer@vu.nl}
\vspace{10pt}
\begin{indented}
\item[] \today
\end{indented}

\begin{abstract}
The helium dimer in its metastable triplet state is a promising candidate to be the first laser-cooled homonuclear molecule. An ultracold gas of \Het would enable a new generation of precision measurements to test quantum electrodynamics for three- and four-electron molecules through Rydberg spectroscopy. 
Nearly diagonal Franck-Condon factors are obtained because the electron employed for optical cycling occupies a Rydberg orbital that does not take part in the chemical bond.
Three possible laser cooling transitions are identified and the spin-rovibronic energy-level structure of the relevant states as well as electronic transition moments, linestrengths, and lifetimes are determined. 
The production of \Het molecules in a supersonic beam is discussed, and a laser slowing scheme to load a magneto-optical trap under such conditions is simulated using a rate equation approach. Various repumping schemes involving one or two upper electronic states are compared to maximize the radiative force. Loss mechanisms such as spin-forbidden transitions, predissociation, and ionization processes are studied and found to not introduce significant challenges for laser cooling and trapping \Het. 
The sensitivity of the vibrational levels of \Hep with respect to the static polarizability of atomic helium is determined and its implications for a new quantum pressure standard are discussed.
\end{abstract}

\section{Introduction}

Precision measurements on few-electron systems are essential for testing and refining the foundations of molecular physics. A prominent example is molecular hydrogen, whose dissociation energy has long served as a benchmark in the development of quantum chemistry and molecular quantum electrodynamics. For this system, agreement between experiment and theory has reached the level of a few parts per billion \cite{holsch2019a}.
Beyond this prototypical two-electron molecule, substantial progress has been achieved in the theoretical description of three- and four-electron systems, notably \Hep\ \cite{tung2012a,matyus2018a,ferenc2020a} and \Hetwo\ \cite{yarkony1989a,komasa2006a,pavanello2008a,cencek2012a}, further extending the scope of \textit{ab initio} methods.
On the experimental side, high-precision measurements of the ionization energy of \Het\ and several spin-rovibrational levels of \Hep\ have been performed using multichannel quantum-defect theory (MQDT) assisted Rydberg extrapolation, yielding agreement with theoretical predictions at the MHz level \cite{semeria2020a, jansen2018a, jansen2018b}. An accurate determination of the rovibrational level structure of the highest vibrational states allows one to extract information about the long-range part of the molecular potential energy curve and the ion--induced-dipole interaction at long range between He and He$^+$ is encoded in the static electric dipole polarizability of atomic helium \cite{semeria2020a}. The polarizability is a critical quantity for the development of new primary pressure standards \cite{gaiser2020a} and can be computed to very high accuracy using quantum electrodynamics (QED) theory \cite{puchalski2016b,puchalski2020b}, but experimental measurements are scarce \cite{gaiser2018a}. 

To date, the accuracy of precision measurements in molecular hydrogen and helium has been fundamentally limited by Doppler-related effects, including both line broadening and systematic shifts arising from the use of supersonic molecular beams.
Although external electric and magnetic field-based slowing techniques have been successfully applied to H$_2$ \cite{seiler2011b} and \Hetwo \cite{motsch2014a}, laser cooling — a prerequisite for substantially increasing phase-space densities — has remained experimentally out of reach.

Direct laser cooling of molecules was first demonstrated for SrF \cite{shuman2010a} and has since resulted in magneto-optical traps (MOTs) for SrF \cite{barry2014a}, CaF \cite{zhelyazkova2014a, truppe2017b, anderegg2017a}, YO \cite{collopy2018a} and BaF \cite{zeng2024a}, even reaching temperatures below the Doppler limit and down to the $\mu K$ range \cite{cheuk2018a,caldwell2019a, ding2020a, wu2021a}. Experiments on several other polar diatomic or polyatomic molecules are underway (see for example Table 1 in \cite{fitch2021a} and \cite{chae2023a}). First attempts to further cool molecules by elastic collisions (evaporative cooling) have consistently observed large losses \cite{gregory2020a,bause2021a,jorapur2024a} caused by sticky collisions, leading to pairs of molecules which are rapidly removed from the trap \cite{mayle2013a,christianen2019a,christianen2019b}. A full quantitative understanding has yet to be achieved, but the loss is related to the large mass and the resulting enormous level density in the collision complexes \cite{christianen2019a}. Elastic collisions for the purpose of evaporative cooling can alternately be induced via collisions with atoms \cite{son2020a,jurgilas2021a}. However, collisional losses still represent a major obstacle to ultracold molecule experiments, as they make evaporative cooling difficult and limit the lifetimes of dense molecular clouds. Recently, Bose-Einstein condensation of NaCs was achieved by using RF-shielding to prevent inelastic collision losses at short range \cite{bigagli2024a}. 

In this work, we explore the prospects for direct laser cooling of the helium dimer (we exclusively consider $^4$He$_2$).
By mitigating the Doppler effect, laser-cooled and trapped \Hetwo molecules could enable a new generation of high-precision measurements on few-electron molecular systems, with accuracies beyond current experimental limits. Moreover, the feasibility of accurate \textit{ab initio} calculations for this light, few-electron system makes cold, controllable \Hetwo an ideal platform for benchmarking theoretical predictions of elastic and inelastic scattering calculations — tests that remain out of reach for heavier, more complex diatomic molecules. Additionally, low-mass systems with a simpler level structure have more favorable collision properties, opening the possibility of collisional cooling to deep quantum degeneracy \cite{son2020a}.

The proposed cooling scheme also expands the class of molecules amenable to direct laser cooling. Existing approaches have primarily focused on species featuring an unpaired electron localized on a metal atom, which does not participate in the chemical bond and yields near-diagonal Franck–Condon factors. In contrast, \Hetwo represents a qualitatively distinct system.

\begin{figure}
    \centering
    \includegraphics[width=0.8\columnwidth]{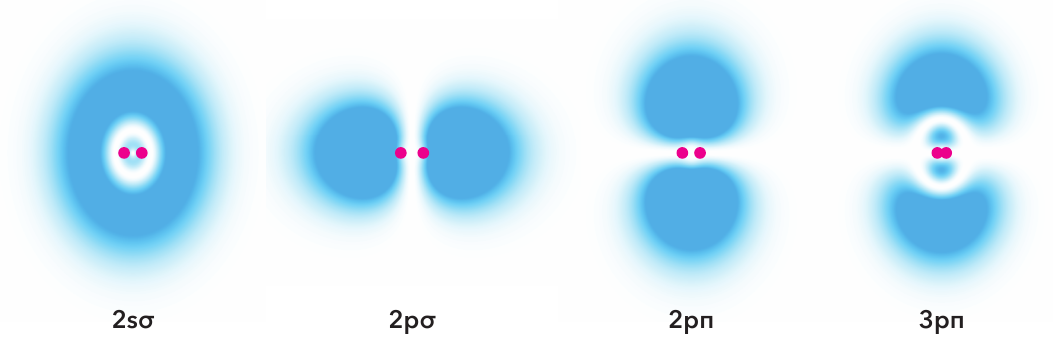}
    \caption{Rydberg electron densities of the lowest triplet Rydberg states of \Hetwo. The nuclei with the equilibrium bond length of $R\approx 2~a_0$ are indicated as dots. }
    \label{fig:ElectronDensity}
\end{figure}

The singlet ground state of the helium dimer, first observed only in 1993 \cite{luo1993b}, is very weakly bound, supporting a single vibrational level. The existence of a long-lived, metastable triplet state, \Het, has been known since 1913 \cite{curtis1913a}.
As early as 1929, Weizel remarked that \Hetwo was perhaps the best-characterized molecule of its time, with more than thirty electronic states identified, serving as an important testbed for the newly developed wave mechanics \cite{weizel1929a}.
The chemical bond in \Hetwo is formed primarily by the three inner electrons, while the outer Rydberg electron occupies a highly excited orbital and contributes little to the chemical bond, as illustrated by the outer electron densities in \Fref{fig:ElectronDensity} for the lowest Rydberg states.

For molecules of this type, that feature strongly bound electronically excited states and a repulsive or weakly bound ground state, Herzberg introduced the term {\it Rydberg molecules} \cite{herzberg1987a}.
Such systems are particularly well suited for optical cycling: excitations predominantly involve the Rydberg electron, minimally perturbing the molecular geometry.  As a consequence, highly diagonal Franck–Condon factors are obtained, and in turn, reduce the number of laser frequencies needed to maintain a closed-loop cooling cycle. 

\Het shares many similarities with the triplet states of molecular hydrogen, however, a notable difference lies in the lifetime of the lowest metastable state. For the $a~^3\Sigma_u^+$ state in \Het, the lifetime has been calculated to be 18\,s \cite{chabalowski1989a}, compared to only a few milliseconds for the metastable $c~^3\Pi_u^-$ state in H$_2$ \cite{astashkevich2015a}. 
This disparity arises from the Wigner–Witmer rules, which predict both singlet and triplet molecular states associated with the lowest dissociation threshold, H(1s) + H(1s), in H$_2$, whereas only a singlet state is permitted at the analogous threshold for He(1s$^2$) + He(1s$^2$). As a result, radiative decay of \Het requires a spin flip, significantly suppressing the transition rate. Additionally, predissociation is also disallowed by spin and inversion symmetry selection rules.
Given typical experimental timescales, the $a$ state of \Het can effectively be regarded as the ground state, making \Het the simplest molecule for which direct laser cooling appears feasible today.
Other examples of Rydberg molecules that may be amenable to optical cycling include heavier rare-gas dimers and simple triatomic species such as H$_3$ \cite{herzberg1987a}.
In the case of heavier rare-gas dimers, the lifetimes of the metastable triplet states are significantly reduced \cite{oka1974a}, primarily due to enhanced spin–orbit coupling, which facilitate radiative decay into the singlet ground state dissociation continuum. Additionally, loss channels arising from predissociation or spin–orbit mixing in the excited states remain key obstacles to establishing closed optical cycling schemes.

Compared to molecules laser-cooled thus far, \Het exhibits several distinct features: it lacks hyperfine structure ($I=0$ for $^4$He) and, owing to the Pauli principle, possesses only half the number of rotational states, which could lead to shorter lifetimes of collision complexes, suppressing losses during evaporative cooling.
Optical cycling in \Het has been employed to detect and image metastable molecules embedded in superfluid helium, where they are produced by ionizing radiation \cite{mckinsey1999a,rellergert2008a}.
It has also been demonstrated that the application of repumping lasers addressing vibrationally excited levels of the metastable state improves cycling efficiency and enhances laser-induced fluorescence.

In the following sections, we review the spin-rovibronic level structure of the helium dimer, identify electronic transitions suitable for laser cooling, and provide detailed information on laser wavelengths, state lifetimes, and branching ratios.
We then discuss key aspects of vibronic and spin-rotational optical cycling, as well as potential loss channels such as predissociation and photoionization.
The paper concludes with an outlook on the prospects of employing laser-cooled \Hetwo for the experimental determination of the dipole polarizability of atomic helium, a quantity needed for the implementation of an SI-tracable quantum pressure standard.

\section{Level structure}

\subsection{Electronic level structure}
As already realized in 1929 by Wentzel, the electronic terms of \Hetwo can be described as Rydberg states $n\ell\lambda$ (indicating the principal quantum number $n$, the orbital angular momentum $\ell$ and its projection along the internuclear axis $\lambda$ of the Rydberg electron) with term values $T_{nl\lambda}$ given by the Rydberg equation
\begin{equation}
    T_{nl\lambda} = T^+(v^+, N^+) - \frac{\mathcal{R}_{\mathrm{He}_2}}{(n-\mu_\ell)^2},
\end{equation}
and with $T^+(v^+, N^+)$ being the rovibrational energy of \Hep, $\mathcal{R}_{\mathrm{He}_2}=109\,729.796 $~cm$^{-1}$ the Rydberg constant, and $\mu_\ell$ the quantum defect. The molecular ion is assumed here to be in the electronic ground state $X^+ \, ^2\Sigma_u^+$ with the electron configuration $(1s\sigma_g)^2(2p\sigma_u)$. The electronic ground state of the ion is strongly bound with a dissociation energy of $D_0=19\,116.116$~cm$^{-1}$ with respect to the He$^+$($^2S$) + He($^1S$) limit and supports 23 vibrational levels \cite{tung2012a}. With an internuclear equilibrium distance of $R_e = 2.042~a_0$ \cite{tung2012a}, the molecular ion has a rotational constant of $B_e\approx 7.22 \,$\wn and a vibrational splitting of $\omega_e\approx 1627.2 \,$\wn \cite{herzberg1950b}. 

\begin{figure}
    \centering
    \includegraphics[width=\columnwidth]{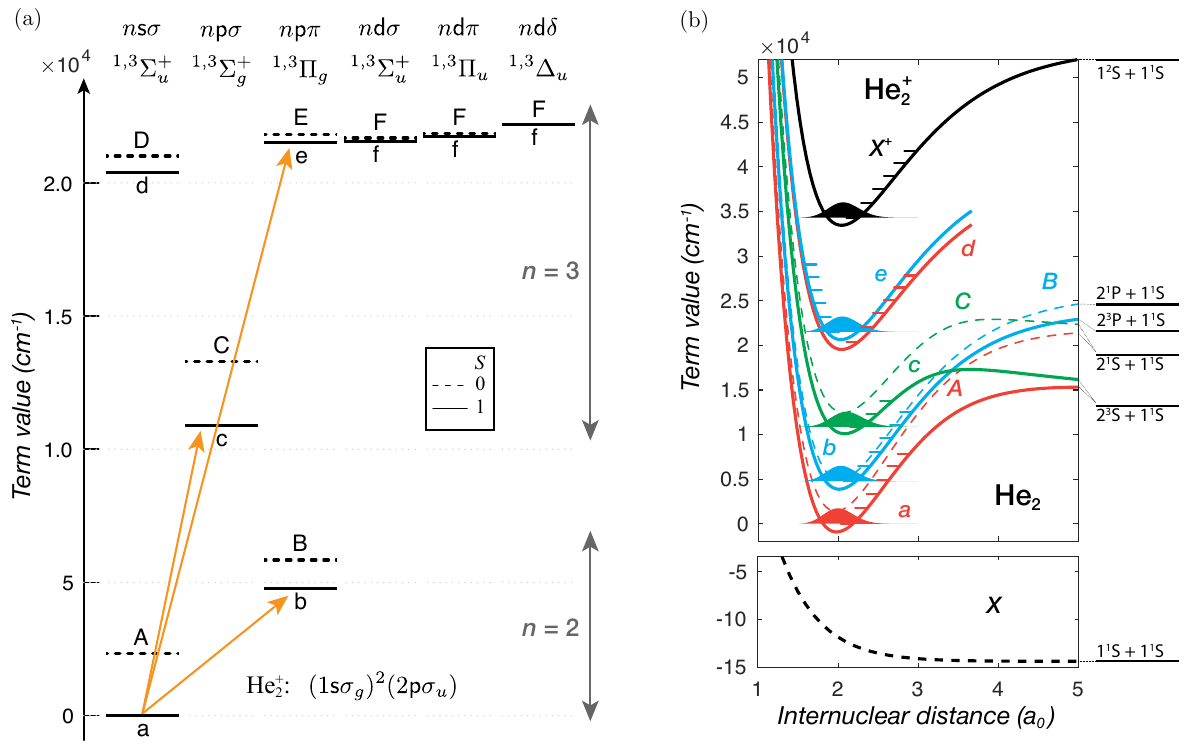}
    \caption{Electronic-level structure of the helium dimer. a) Energy-level diagram of the $n=2$ and 3 singlet $(S=0)$ and triplet $(S=1)$ Rydberg states with $\ell=0,1,2$. b) Potential-energy curves of the lowest singlet (dashed line) and triplet (solid line) Rydberg states of \Hetwo. The ground state of the neutral and ion dimer are indicated in black and atomic dissociation products are indicated. The nuclear wave function for the vibrational ground state of the triplet states and the ionic ground state are shown. }
    \label{fig:PES}
\end{figure}

The addition of the Rydberg electron to the strongly bound ion core gives rise to strongly bound Rydberg states, each supporting several vibrational levels. Given the open-shell ion core with $S^+=1/2$, singlet and triplet Rydberg states can be obtained. Figure~\ref{fig:PES} (a) indicates the experimental term values of all possible singlet (dashed lines) and triplet (solid lines) Rydberg states with principal quantum numbers $n=2$ and 3 \cite{herzberg1950b}. The term values are given with respect to the lowest triplet ground state $a~2s\sigma$, which is metastable. The singlet-triplet splitting caused by the exchange interaction between core and Rydberg electrons decreases with increasing $n$ and increasing $\ell$ as expected, and already the 3d states seem degenerate on the scale of Figure~\ref{fig:PES}.   

The $2p\sigma$ Rydberg state is missing, as this state corresponds in the united atom description to the repulsive electronic ground state $X~^1\Sigma_g^+$ of \Het with the electron configuration $(1s\sigma_g)^2(2p\sigma_u)^2$, dissociating into He($^1S$) + He($^1S$). 
Figure~\ref{fig:PES} (b) displays the potential energy curves of the electronic ground states of the ion \cite{tung2012a} and the neutral molecule \cite{yarkony1989a}, respectively, as well as the lowest Rydberg states \cite{yarkony1989a}. The defining characteristics of a \emph{Rydberg molecule} are clearly visible: a repulsive ground state of the neutral molecule and strongly bound excited states with potential energy curves $U_\mathrm{BO}(R)$ which resemble the one of the molecular ion $U^+_\mathrm{BO}(R)$, but are shifted by the binding energy of the Rydberg electron $\epsilon(R)$
\begin{equation}
    U_\mathrm{BO}(R) = U^+_\mathrm{BO}(R) + \epsilon(R) = U^+_\mathrm{BO}(R) - \frac{1}{2(n-\mu_\ell(R))^2}.
\label{eq:PEC}
\end{equation}
The vanishing $R$-dependence of the quantum defect $\mu_\ell(R)$ can be illustrated by comparing the rotational (7.66\wn, 7.46\wn, 7.00\wn, 7.34\wn and 7.32\wn) and vibrational (1811.2\wn, 1697.7\wn, 1481.0\wn, 1654.0\wn and 1724.6\wn) constants of the $a, b, c, d$ and $e$ states \cite{herzberg1950b} with the ones of the ion, indicating that the minimum and shape of the potential are only changing marginally.  

Notable outliers in Figure~\ref{fig:PES} (a) are the $c$ and $C$ state, which start to deviate from the ionic curve at around $3a_0$, forming a hump, before dissociating to the He(2s) + He limit. Mulliken analyzed the dissociation behavior in detail \cite{mulliken1970a} and could trace back the appearance of those humps to avoided crossings with repulsive Rydberg states with the ion core in the $A^+ \, ^2\Sigma_g^+$ (repulsive) excited state with electron configuration $(1s\sigma_g)(2p\sigma_u)^2$. In H$_2$ a similar mechanism leads to the formation of double-well states like EF and GK, by a first avoided crossing at small $R$ between Rydberg states with the same symmetry but different ion core configuration and a second avoided crossing at large $R$ with the ion pair configuration H$^+$H$^-$. With He$^+$He$^-$ inaccessible at this energy range, the second avoided crossing does not take place in \Het, leading to the aforementioned humps instead. 

\subsection{Spin-rotational level structure}
The spin-rotational level structure in the lowest Rydberg states of \Het is best described in Hund's case (b) for both $\Sigma$ and $\Pi$ states, given that $A/B \approx 0.03 \ll 1$, where $A$ and $B$ are the molecular constants for the spin-orbit and rotation energies, respectively. For the triplet states, each rotational level is split into three components: $F_1$ ($J=N+1$), $F_2$ ($J=N$) and $F_3$ ($J=N-1$). In addition, one must evaluate restrictions on the values of the rotational angular momentum $N$ due to the generalized Pauli principle. The behavior of the total molecular wave function $\psi_\mathrm{tot} = \psi_\mathrm{e} \cdot \psi_\mathrm{vib} \cdot \psi_\mathrm{rot} $ under permutation (12) of the two alpha particles is given as $(-1)^{t_1}(-1)^{t_2}(-1)^N$, where $t_1$ distinguishes between \emph{gerade} ($t_1=0$) and \emph{ungerade} ($t_1=1$) states and $t_2$ between $\Sigma^+, \Pi^+, ...$ ($t_2=0$) and $\Sigma^-, \Pi^-, ...$ ($t_2=1$). The levels are labeled as symmetric ``s" and antisymmetric ``a", respectively, and we note that, for the case of two indistinguishable bosonic alpha particles, only ``s" levels are allowed. 

\begin{figure}
    \centering
    \includegraphics[width=0.6\columnwidth]{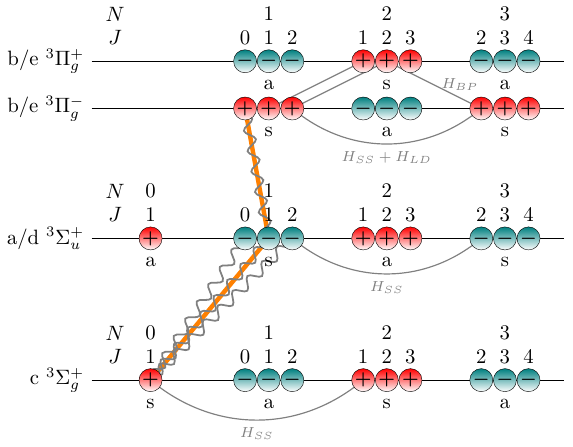}
    \caption{Symmetry properties of spin-rotational levels for \Hetwo. Level interactions caused by off-diagonal elements in the Hamiltonian are indicated, where $H_{BP} = H_{SO} + H_{SS} + H_{LD}$. Laser excitation (orange line) and the dominant fluorescence pathways (gray wavy lines) for the laser cooling transitions are shown. }
    \label{fig:Symmetry}
\end{figure}

\Fref{fig:Symmetry} indicates the permutation symmetry and the total parity $E^*$ for $^3\Sigma_u^+$, $^3\Sigma_g^+$ and $^3\Pi$ states (for singlet states the permutation and parity symmetry is the same, because the symmetry is not affected by the value of $S$). The nuclear-spin statistics theorem eliminates all of the even rotational levels (+ parity) for the $a$ and $d$ state and all odd rotational levels ($-$ parity) of the $c$ state. For the $np\pi$ states $b$ and $e$, half of the $\Lambda$-doubling components are eliminated and only the levels with (+) parity remain.  

The effective rotational Hamiltonian can be written in Hund's case (b) basis as
\begin{eqnarray}
    \hat{H}_{\mathrm{eff}} & = \hat{H}_R + \hat{H}_{SR} + \hat{H}_{SS} + \hat{H}_{SO} + \hat{H}_{LD} = \nonumber \\
    & = B \, \hat{\textbf{N}}^2 +  \gamma \,\hat{\textbf{N}} \cdot \hat{\textbf{S}} + \frac{2}{3}\lambda \, (3\hat{S}_z - \hat{\textbf{S}}^2) + A_{SO} \, \hat{L}_z \hat{S}_z +  \frac{q}{2}(\hat{N}^2_+  e^{-2i\phi} + \hat{N}^2_- e^{2i\phi}) + \nonumber \\
    & + \frac{o}{2}(\hat{S}^2_+  e^{-2i\phi} + \hat{S}^2_- e^{2i\phi}) - \frac{p}{2}(\hat{N}_+\hat{S}_+  e^{-2i\phi} + \hat{N}_-\hat{S}_- e^{2i\phi}).
    \label{eq:EffectiveHamiltonian}
\end{eqnarray}
In the last three terms, which correspond to the $\Lambda-$doubling Hamiltonian, $\phi$ is the electron orbital azimutal angle, and the presence of $e^{\pm i2\phi}$ ensures that only matrix elements between $\bra{\Lambda = \pm1}$ and $\ket{\Lambda = \mp 1}$ are non-zero \cite{brown2003a}. For $\Sigma$ states, only the first three terms in \eqref{eq:EffectiveHamiltonian} appear. 

\begin{figure}
    \centering
    \includegraphics[width=\columnwidth]{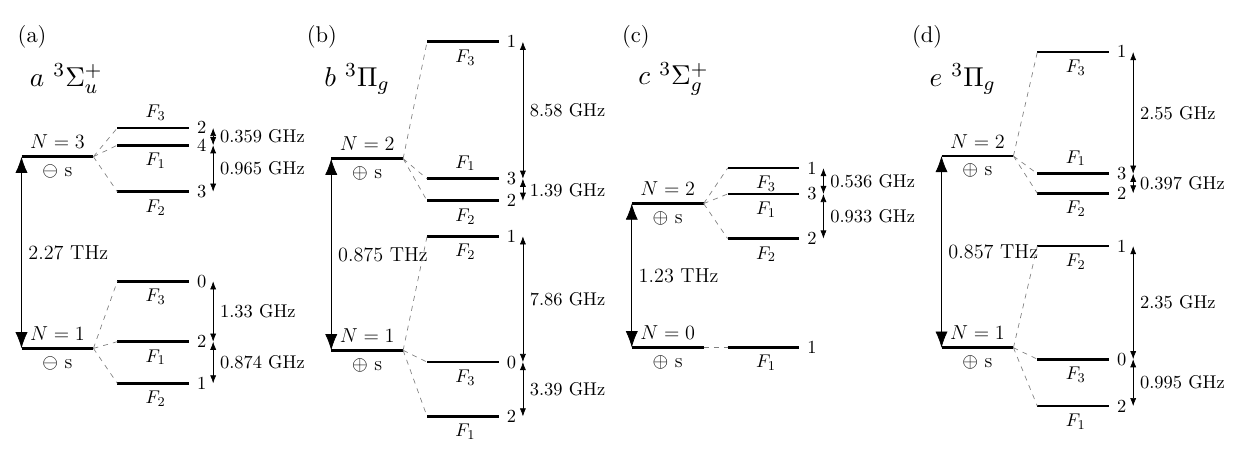}
    \caption{Fine structure of the lowest rotational levels of the $a$, $b$, $c$ and $e$ states. }
    \label{fig:FineStructure}
\end{figure}

The rotational energy of each electronic state can be computed by diagonalizing the effective Hamiltonian using the values of the molecular constants given in \Tref{tab:MolecularConstants}. Although some terms in $\hat{H}_{\mathrm{eff}}$ introduce weak off-diagonal couplings between rotational states (see \Fref{fig:Interactions}), the level structure is still well approximated by Hund’s case (b). \Fref{fig:FineStructure} shows the fine structure of the first rotational levels of $a$, $b$, $c$ and $e$ state. Notable differences exist between $\Sigma$ and $\Pi$ states regarding the ordering and splitting of the fine-structure components.

In the case of $\Sigma$ states, the fine structure splitting is caused by spin-rotational and spin-spin interactions. According to \Tref{tab:MolecularConstants}, the relation $\hat{H}_{SS} \gg \hat{H}_{SR}$ is fulfilled for $a$, $c$ and $d$, which implies that the largest contribution to the splitting between the fine structure levels comes from $\hat{H}_{SS}$. In addition, the large separation between the rotational states leads to vanishing off-diagonal elements of $\hat{H}_{SS}$. Assuming no spin-rotational contribution and considering only the diagonal matrix elements, the energy of a fine structure level with rotational angular momentum $N$ and total angular momentum $J$ is given by

\begin{eqnarray*}
    J = N+1: \qquad  & \frac{2}{3}|\lambda|  \, \, \frac{N+1}{2N-1}, \\
    J=N: \qquad & -\frac{2}{3}|\lambda|, \\
    J=N-1: \qquad & \frac{2}{3}|\lambda|   \, \, \frac{N}{2N+3} ;
\end{eqnarray*}
where we have used that $\lambda < 0$. Since $$\frac{N+1}{2N-1} > \frac{N}{2N+3} > -1$$ for any value of $N$, the energy levels always follow the order $F_3 > F_1 > F_2$. 

For $\Pi$ states, the order of the fine structure levels changes for different values of $N$. In this case, the dominant constants are $o \approx A_{SO} > \lambda \gg \gamma$. The first-order contribution $o^{(1)}$ comes from the spin-spin interaction. Taking again the diagonal elements, one finds that the term associated to $o^{(1)}$ depends on $(-1)^N$, which leads to a change in the order for even and odd rotational quantum numbers. For low values of $N$, $\braket{\hat{H}_{SO}}$ is stronger than $\braket{\hat{H}_{SS}}$, giving the following order:

\begin{eqnarray*}
    & F_3 > F_1 > F_2 \qquad \qquad \mathrm{for~} N~\mathrm{ even}, \\
    & F_2> F_3 > F_1 \qquad \qquad \mathrm{for~} N~\mathrm{ odd}.
\end{eqnarray*}

\begin{table} \footnotesize
\caption{\label{tabone} Molecular constants in the effective Hamiltonian (cm$^{-1}$). Uncertainties are given as subscripts.} 
\lineup
\begin{tabular}{@{}*{8}{l}}
\br
&$a \,^3 \Sigma_u^+$&$b \, ^3\Pi_g$&$c \, ^3\Sigma_g^+$&$d \, ^3\Sigma_u^+$&$e \, ^3\Pi_g$&$X^+$$^2\Sigma^+_u$\\
\mr
$T$& 0 & 4768.14542$_{35}$ &10889.4717$_{19}$ & 20391.3 & 21507.25$_1$ & 34301.20700$_4$\\
$\omega_e$ & 1808.5$_{6}$ &1769.07$_2$ & 1583.85 & 1728.01$_2$ & 1721.1$_9$ & 1628.3832$_{12}$\\
$B$ & \m$7.589141_{27}$ & \m$7.323430_{29}$ & \m$6.853952_{40}$ & \m$7.228_{1}$&\m$7.1728_{8}$&\m$7.101143_{25}$\\
$\gamma$\tiny{$\, \times 10^{5}$} & $-8.0805_{22}$ & $-11.93_{129}$ & $-8.0805$ & $-8.0805$ & $-11.93$&$-16.8_3$\\
$\lambda$\tiny{$\, \times 10^{2}$}&$-3.6664342_{128}$&\m$5.5554_{30}$&$-3.6664342$&$-1.086$&\m$1.6456$&--- \\
$A_{SO}$&--- &$-0.22733_{82}$&--- &--- &$-0.06736$&--- \\
$o$&--- &$-0.28975_{37}$&--- &--- &$-0.085852$&--- \\
$p$\tiny{$\, \times 10^{4}$}&--- &\m$5.59_{24}$&--- &--- &\m$5.59$&--- \\
$q$\tiny{$\, \times 10^{2}$}&--- &$-2.53917_{94}$&--- &--- &$-2.53917$&--- \\
Ref.& \cite{focsa1998a} & \cite{focsa1998a} & \cite{focsa1998a} & \cite{ginter1965e} & \cite{brown1971b} & \cite{semeria2020a} \cite{jansen2018a} \cite{semeria2016a} \\
\br
\end{tabular}
\label{tab:MolecularConstants}
\end{table}

Since for the $d$ and $e$ states there is no experimental data on the fine structure available, the respective scaled parameters of the $a$ and $b$ states were used, belonging to the $ns\sigma_u$ and $np\pi_g$ Rydberg series, respectively. Since the spin-spin and the spin-orbit interactions depend on $r^{-3}$, the parameters $A_{SO}$, $\lambda$ and $o$ can be scaled as $n^{-3}$ for the same Rydberg series. In contrast, the spin-rotation stems mainly from the core electrons, so $\gamma$ is assumed to be $n$ independent \cite{jansen2018a}.

\section{Radiative lifetimes and branching ratios}

\subsection{Line strengths}
The matrix element for an electric dipole transition connecting a lower state $\ket{i}=\ket{\eta,\Lambda;v;J\Lambda M}$ to an upper state $\bra{f}=\bra{\eta',\Lambda';v';J'\Lambda' M'}$ is given by
\begin{eqnarray}
  \label{eq:WET}
  \Braket{f | -\mathbf{T}^{(1)}(\varepsilon) \cdot \mathbf{T}^{(1)}(\mu) | i} 
  &=& -\sum_{p,q} (-1)^p 
  \Braket{f | T^{(1)}_{-p}(\varepsilon) 
  \mathcal{D}^{(1)*}_{p\,q}T^{(1)}_q(\mu) | i} \nonumber \\
  &&\hspace*{-17em} =-\sum_{p,q} (-1)^p T^{(1)}_{-p}(\varepsilon) 
  (-1)^{J'-M'} 
  \Wthreej{J'}{1}{J}{-M'}{p}{M} \nonumber \\
  &&\hspace*{-5em} \times 
 \Braket{\eta',\Lambda';v';J'\Lambda' || 
  \mathcal{D}^{(1)*}_{.\,q}T^{(1)}_q(\mu) || 
  \eta,\Lambda;v;J\Lambda}
\end{eqnarray}
The line strength $S$ is defined by summing \eqref{eq:WET} over all possible polarizations $p$, $M'$ and $M$, and can be seen to correspond to the reduced matrix element, $\braket{||\ldots||}$, when taking into account the orthogonality of the Wigner 3j-symbol. 
The expression for $S$ can be further simplified
\begin{eqnarray}
  \label{eq:linestrengthM2S}
  S_{fi} &=& \Braket{\eta',\Lambda';v';J'\Lambda' || \mathcal{D}^{(1)*}_{.\,q}T^{(1)}_q(\mu) || \eta,\Lambda;v;J\Lambda}^2 \nonumber \\
  &=& \Braket{\eta',\Lambda';v' || T^{(1)}_{q}(\mu) || \eta,\Lambda;v}^2 
  \Braket{J'\Lambda' M' || \mathcal{D}^{(1)*}_{.q} || J\Lambda M }^2 \nonumber \\
  &=& \mathscr{M}^2_{\eta'\Lambda'v',\eta\Lambda v} 
  \mathscr{S}_{\Lambda' N'SJ',\Lambda NSJ}
  \end{eqnarray}
where $\mathscr{M}$ is the vibronic transition moment and $\mathscr{S}$ is the H\"onl-London factor. For parallel transitions $\Delta\Lambda=q=0$ and for perpendicular transitions one finds $\Delta\Lambda=q=\pm1$.

The vibronic transition moment is evaluated using the vibrational wave functions $\psi_v=\braket{R|v}$ and the electronic transition moments $\mu_{\eta'\Lambda',\eta\Lambda}(R)$, taken from the literature or evaluated using the Coulomb approximation (see Appendix A).

For the angular factor, the reduced matrix element is obtained by uncoupling the electron spin from $J=N+S$ and then evaluating the matrix elements of the Wigner rotation matrix, resulting in
\begin{eqnarray}
  \label{eq:linestrength}
  \mathscr{S}_{\Lambda N'SJ',\Lambda NSJ} 
  &=& (2J'+1)(2J+1) \Wsixj{N}{J}{S}{J'}{N'}{1}^2 \nonumber \\
  &&\times\,(2N'+1)(2N+1) \Wthreej{N'}{1}{N}{-\Lambda'}{q}{\Lambda}^2
  \end{eqnarray}
where the expression in the second line can be identified as the H\"onl-London factor neglecting spin, $\mathscr{S}_{\Lambda' N',\Lambda N}$. When summing over all fine-structure components in the initial and final rotational state, $\sum_{J'J }\mathscr{S}_{\Lambda N'SJ',\Lambda NSJ} = (2S+1)\mathscr{S}_{\Lambda' N',\Lambda N}$.

\begin{figure}
    \centering
    \includegraphics[width=\columnwidth]{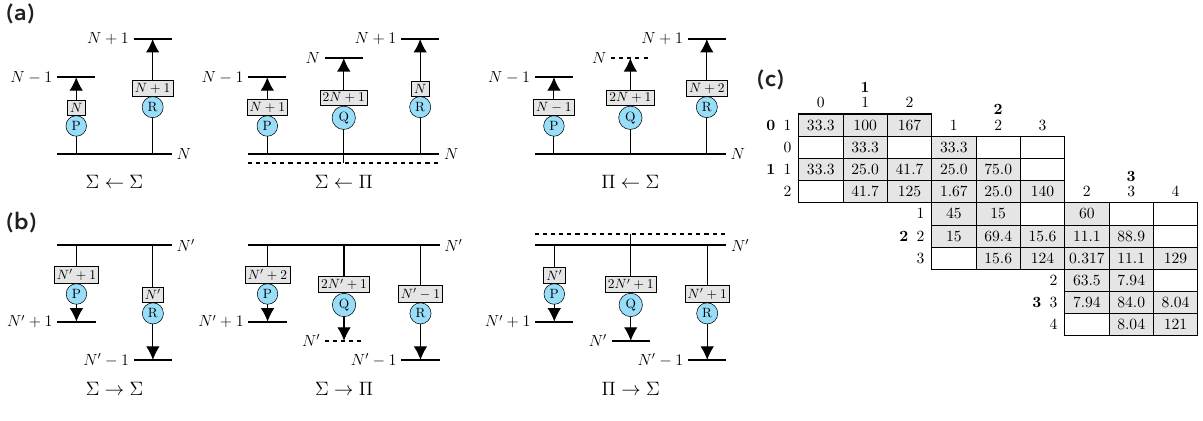}
    \caption{ Line strength factors for (a) absorption and (b) emission. $\Lambda$-doublets are distinguished when needed by a dashed line. For fine-structure transitions the spin-part of the line strength is multiplied by 100 and given in (c). }
    \label{fig:linestrength}
\end{figure}

\Fref{fig:linestrength} summarizes the rotational line strengths for absorption and emission for $\Sigma-\Sigma$ and $\Sigma-\Pi$ transitions in terms of the rotational quantum number of the ground ($N$) and excited ($N'$) state, respectively.  
The H\"onl-London factor for a specific fine-structure component (multiplied by 100) is shown in \Fref{fig:linestrength} (c), where each block for a given pair of $N$ and $N'$ adds up to 300.

The Einstein coefficient for the emission from $\ket{f}$ to $\ket{i}$ is defined as \cite{larsson1983a, whiting1972a}
\begin{equation}
    A_{fi} = \frac{64\pi^4\tilde{\nu}^3}{(4\pi\epsilon_0)3h} e^2a_0^2 |\Braket{f | r | i}|^2,
\end{equation}
where $\tilde{\nu}$ is the frequency of the transition in cm$^{-1}$ and $\Braket{f | r | i}$ is given in atomic units. The numerical conversion factor amounts to $2.026\times10^{-6}$ with the Einstein A coefficient in s$^{-1}$. The total decay rate for a transition $J'\to J$ is obtained by summing over all $M',M$ and $p$ to relate it to the line strength $S_{fi}$ and divide by the degeneracy factor of the upper level
\begin{equation}
    A_{\Lambda'v'N'J',\Lambda vNJ} = \frac{64\pi^4\tilde{\nu}^3}{(4\pi\epsilon_0)3h} e^2a_0^2 \frac{S_{\Lambda'v'N'J',\Lambda vNJ}}{2J'+1}.
\end{equation}

The total decay rate for a fine-structure component $J'$ is obtained as a sum over all possible $(\Lambda,v,N,J)$.

\subsection{Branching ratios and lifetimes}

The branching ratio $b_{fi}$ is given by,
\begin{equation}
    b_{fi} = \frac{A_{fi}}{\sum\limits_{j} A_{fj}} = A_{fi}\tau_f ,
\end{equation}
where the sum in the denominator is over all possible lower states to which $\ket{f}$ can decay and $\tau_f$ is the radiative lifetime. The lifetimes and the electronic branching ratios of the triplet excited states can be found in \Tref{tab:Lifetimes}. 

The long lifetime of the $b$ state is a result of its proximity in energy to the $a$ state and the implications for its use in a cooling scheme will be discussed below. The other excited states exhibit comparable lifetimes, arising from a balance between the increasing transition frequency and the decreasing electronic transition moment with increasing $\Delta n$ ($\mu_{ac} \approx 3 \mu_{ae}$, see \ref{sec:ElectronicTransitionMoments}). 

It is apparent from \Fref{fig:PES} that $e$ and $d$ have different decay paths, motivating a study of their respective electronic branching ratios. The large transition energy makes the decay from $e$ to $a$ dominant over the decay to $d$. The effect of the 0.7\% branching ratio to $d$ on the optical cycling process will be discussed below. For the branching ratios $b_{de}$ and $b_{dc}$, the frequency difference and the additional factor of 2 that appears in the line strength for $\Sigma \rightarrow \Pi$ transitions (see \Fref{fig:linestrength}b ) compensate for the smaller electronic transition moment of $b$, favoring $d \rightarrow b$ over $d\rightarrow c$.

Experimental information on lifetimes and branching ratios is scarce. In Ref. \cite{neeser1994a}, the lifetimes for the $e$ and $d$ states were determined in a gas cell at low pressure to be $(57 \pm 10) \,$ns and $(25 \pm 5) \,$ns, respectively. A comparison with \Tref{tab:Lifetimes} shows good agreement for the $e$ state, but a longer theoretical lifetime for the $d$ state. This could originate from unaccounted collisional quenching in the experiment if there is a significant $\Lambda$-dependence in the collision cross sections. The $d$ state lifetime was also measured in liquid helium with a result of $(48 \pm 2) \,$ns \cite{rellergert2008a}.

\begin{table} \caption{\label{label}Radiative lifetimes and electronic branching ratios of excited states in \Het.} 
\begin{indented} 
\lineup
\item[]
\begin{tabular}{@{}llll}
\br
State& $\tau$ (ns) & \multicolumn{2}{r}{branching ratio}    \\
\mr
\textit{e}    &   \055.89      &   0.993~($a$)   & 0.007~($d$)  \\
\textit{d}    &   \052.68      &   0.567~($b$)   & 0.433~($c$)   \\
\textit{c}    &   \044.76      &   1~($a$)  &   \\
\textit{b}    &   642.9       &   1~($a$)  &   \\
\br
\end{tabular}
\end{indented}
\label{tab:Lifetimes}
\end{table}

\section{Production of \Het molecules}
Metastable states of \Het are readily produced in both, continuous and high repetition-rate pulsed beam sources, overcoming limitations of existing molecular MOT experiments \cite{miller1979a,motsch2014a}. The discharge excites ground state He$(1~^1S)$ and creates metastable He$^*(2~^3S)$ atoms, which form metastable \Het molecules via He$(1~^1S)$ + He$^*(2~^3S)$~$\to$~\Het. Electronic excitation in the molecule resulting from the discharge is quenched on a fast time scale (see the discussion in the previous section), with molecules accumulating exclusively in the metastable ground state $a~^3\Sigma_u^+$. 

The molecular beam velocity can be reduced to around 530~m/s and a translational temperature of 1.8~K by keeping the expansion valve at 10~K \cite{motsch2014a}. The light mass and the resulting large rotational and vibrational splittings, avoids the population of a large range of rovibrational levels. An estimated 10\% of molecules can be found in the $a~^3\Sigma_u^+(v=0,N=1)$ rovibrational ground state level and optical pumping can be used to brighten the beam and increase the number of rovibrational ground state molecules \cite{bahns1996a}. 

An Even-Lavie valve with dielectric-barrier discharge was shown to produce $10^{10}$ metastable He atoms per 20$\, \mu$s pulse in a skimmed, supersonic beam at room temperature. With discharge conditions optimized for the production of \Het, and the valve operating at 10~K \cite{motsch2014a}, the production of $10^9$ rovibrational ground state molecules per pulse seems feasible. This number is similar to currently used buffer-gas beam sources, but with a much shorter pulse length and less velocity spread ($\sim2$\,K vs. $\sim13$\,K) \cite{truppe2017b}. With repetition rates up to 1500\,Hz \cite{barnea2021a}, a 20$\, \mu$s \Het pulse could be produced every 667$\, \mu$s, so that more than 150 pulses can be accumulated in a MOT with 100~ms lifetime.   

The high initial forward velocity inherent to supersonic molecular beams presents a challenge for direct laser slowing. While alternative deceleration techniques such as Zeeman deceleration \cite{motsch2014a} and Zeeman–Sisyphus decelerator \cite{augenbraun2021a} have been explored, buffer-gas cooling may also offer a viable approach. For collisions between \Het\ and ground-state He ($^1S_0$) there exists a substantial energy mismatch that should suppress efficient collisional de-excitation (see \Fref{fig:PES}).

\section{Optical cycling}

\begin{table} \caption{\label{tab:cooling_trans}Relevant parameters for laser cooling transitions in He$_2$. See text for details.}  
\lineup
\footnotesize
\begin{tabular}{@{}llllllll}
\br
Transition& $\lambda_{00}$ (nm) &  $\Gamma/2\pi$ (MHz) &
$v_\mathrm{rec}$ (cm/s) & 
$a_\mathrm{max}\,$(m/s$^2$) &
$I_\mathrm{sat}$~(mW/cm$^2$) & $T_\mathrm{Doppler}$ (K) & $T_\mathrm{rec}$ (K) \\
\mr
\textit{a-b}    &   2097.3      &   2.48[-1]   & \02.38  &  1.86[4]    & 3.53[-3]  & 6.0[-6]   & 5.4[-7]\\
\textit{a-e}    &   \0464.6     &   2.85   & \01.07[1]  &  9.66[5]     & 3.71      & 6.8[-5]   & 1.1[-5]\\
\textit{a-c}    &   \0918.3     &     3.56  &  \05.43  &  6.47[5]   & 6.41[-1]     & 9.1[-5]   & 2.8[-6]\\
\br
\end{tabular}
\end{table}

\Tref{tab:cooling_trans} summarizes key parameters for the three transitions considered for optical cycling in \Hetwo, including the main cooling wavelength $\lambda_{00}$, fluorescence decay rate $\Gamma$, the saturation intensity $I_\mathrm{sat}=\pi hc\Gamma/(3\lambda^3)$ and the recoil velocity $v_{\mathrm{rec}} = \hbar k/m$, with $k=2\pi/\lambda_{00}$ being the wave vector and $m=8.00520650826(12) \,$u  the molecular mass \cite{coursey2013a}. Additionally, the maximum radiative acceleration for a two-level system, $a_\mathrm{max}=\hbar k \Gamma/2m$, is given.

The long wavelength and correspondingly small decay rate of the $a–b$ transition render it unsuitable for the initial slowing stage. Nevertheless, its low Doppler ($k_B \, T_{\mathrm{Doppler}} = \hbar\Gamma/2$) and recoil ($k_B \, T_{\mathrm{rec}} = \hbar^2 k^2/m$) temperatures suggest that it may be advantageous for narrow-line cooling applications \cite{stellmer2013a, saskin2019a,mehling2025a}.
In contrast, the $a–c$ and $a–e$ transitions exhibit similar decay rates; however, the $a–e$ transition is associated with a recoil velocity approximately twice that of the $a–c$ transition. This larger recoil favors its use for efficient laser slowing of the supersonic molecular beam.

\subsection{Vibronic optical cycling} \label{sec:Vibronic}

\begin{figure}
    \centering
    \includegraphics[width=\columnwidth]{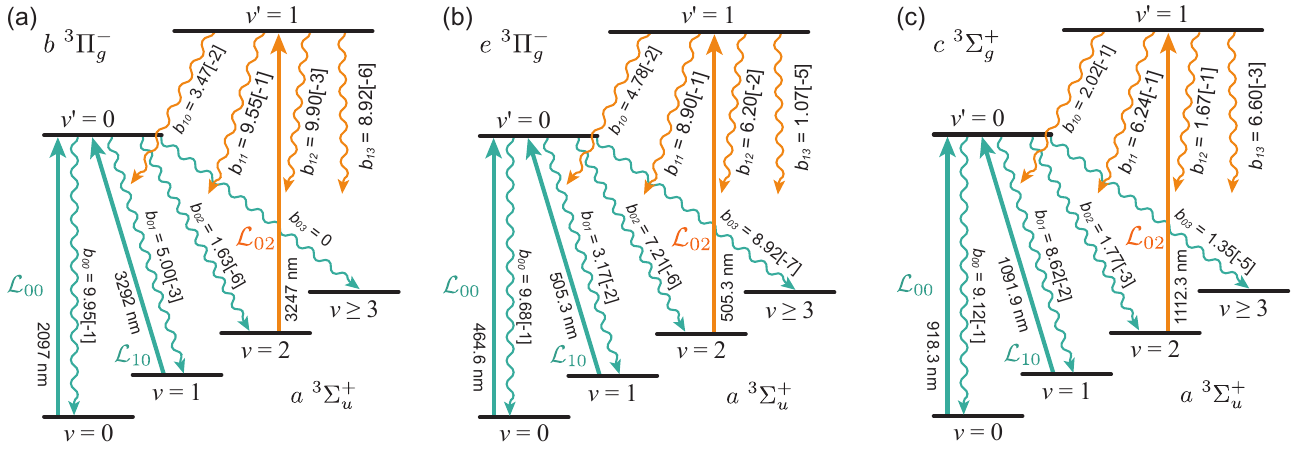}
    \caption{Vibrational branching ratios for (a) $a-b$, (b) $a-e$ and (c) $a-c$ laser cooling transitions. Branching ratios and laser wavelengths for selected transitions are indicated.}
    \label{fig:BranchingRatios}
\end{figure}

The vibronic transition moments $\mathscr{M}^2_{\eta'\Lambda'v',\eta\Lambda v}$ in \eqref{eq:linestrengthM2S} were obtained using the $R-$dependent electronic transition moments for $a-b$ and $a-c$ given in \cite{yarkony1989a} and the $R-$independent transition moment for $a-e$ obtained via the Coulomb approximation. When the $R-$dependence is ignored, $\mathscr{M}^2_{\eta'\Lambda'v',\eta\Lambda v}=\mu_{ij}^2FC$, where FC is the Franck-Condon factor. The use of the FC approximation changes the branching ratios in \Fref{fig:BranchingRatios} on the order of $10^{-3}$, indicating that the electronic character changes only slightly when changing the internuclear distance. 

The results are shown in \Fref{fig:BranchingRatios} for the three cooling transitions (see also \Tref{tab:VibBr} in \ref{Ap:VibBr}). For similar FC factors, light molecules will show smaller off-diagonal branching ratios given their larger vibrational splittings and the $\nu^3$ factor in the Einstein A coefficient.

The best vibrational branching ratio, $>99\%$, is found for the $a-b$ cycling transition, for which only the $v=0,1,2$ levels of the $a$ state can be populated by fluorescence from $b(v'=0)$. Using only two vibrational repump lasers would allow to completely close the vibrational cooling cycle. For the $a-e$ transition, exciting the next Rydberg state in the $np\pi$ series, the branching ratio reduces to $97\%$ for the main cooling transition and is approaching the limiting value corresponding to the $a-X^+$ FC factor. The worst branching ratio, $91\%$, is obtained for $a-c$, as the excited state potential curve is perturbed by the avoided crossing with a core-excited Rydberg state. Compared to the $a$–$np\pi$ transitions, \Fref{fig:BranchingRatios} shows that the off-diagonal branching ratios for the $a$–$c$ transition decrease more slowly with increasing $\Delta v$. This feature allows the $c$ state to serve as a gateway for accessing higher vibrational levels in the ion, but it also complicates the closure of the optical cycling loop.    

\begin{figure}
    \centering
    \includegraphics[width=\columnwidth]{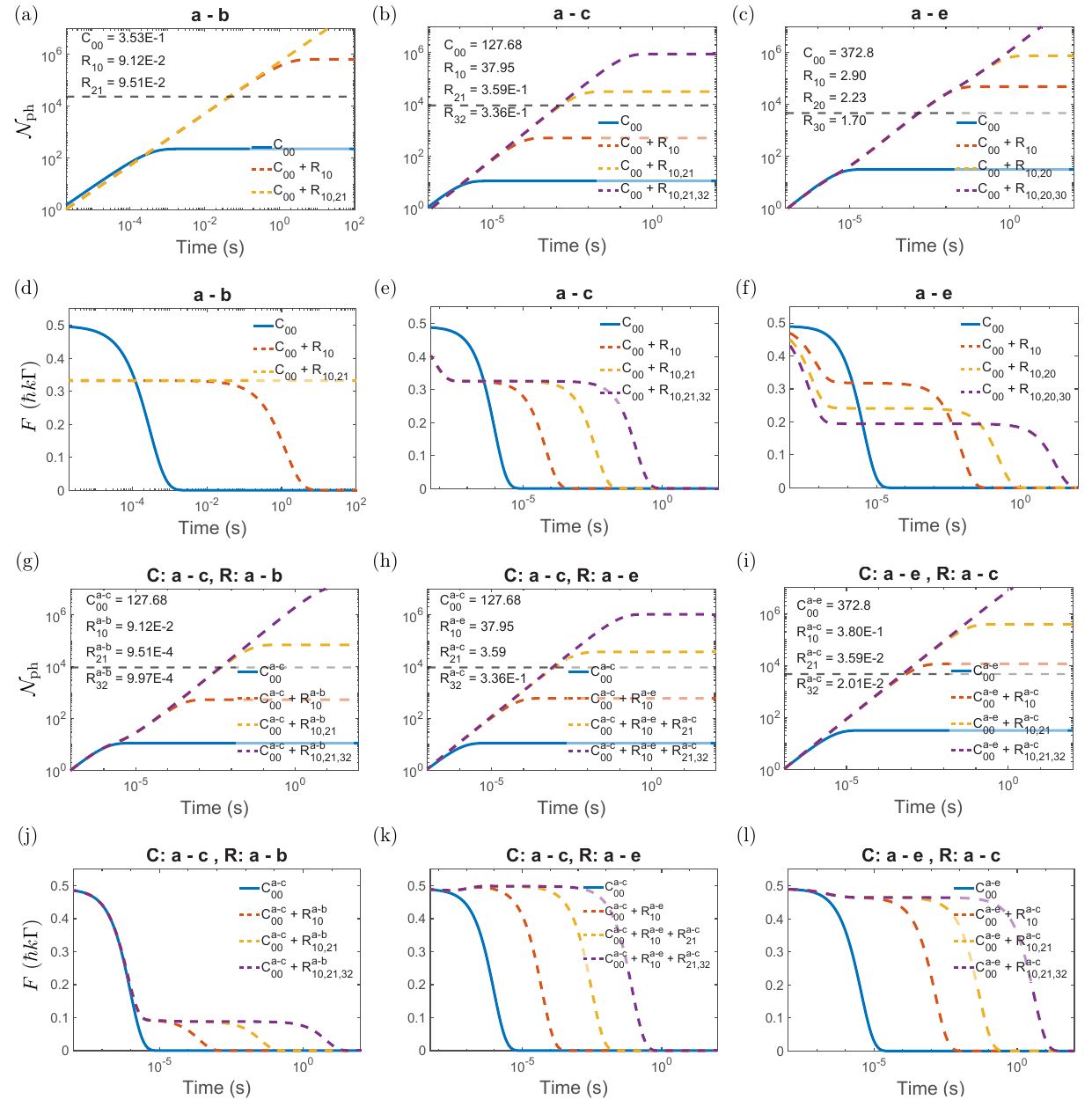}
    \caption{
    The number of scattered photons and radiative force as a function of time for various repump schemes. C$_{if}$ and R$_{if}$ indicate laser-driven transitions used for cooling and repumping, respectively, between vibrational states \textit{i} and \textit{f}. The intensities of both cooling and repump lasers are optimized to achieve sufficient scattering rates, as specified in the legend (in mW/cm$^2$). The radiative force is in units of $\hbar k\Gamma$ for each cooling transition. \textbf{(a)-(f)} illustrate repump schemes with only one upper electronic state. \textbf{(g)-(l)} illustrate repump schemes involving a second electronic state. The gray dashed line indicates the required number of scattered photons, blue solid line and colored dashed lines indicate the laser scheme shown in the legend.}
    \label{fig:OpticalCycling}
\end{figure}

The deceleration of the molecules and the number of scattered photons $\mathcal{N}_{\mathrm{ph}}$ is modeled using a multilevel rate-equation approach for the populations $N$ \cite{tarbutt2015a}:
\setcounter{equation}{9} 
\begin{eqnarray}
\label{rateeq_a}
\frac{dN_i}{dt} &=& \sum_{i,p}R_{fi,p}(N_f - N_i) + \Gamma \sum_i b_{fi} N_u, \\
\label{rateeq_b}
\frac{dN_f}{dt} &=& -\Gamma N_f + \sum_{i,p} R_{fi,p}(N_i - N_f), \\
\label{rateeq_c}
\vec{F} &=& m\hbar \sum_{i,f,p} \vec{k}_{fi,p} R_{fi,p}(N_i - N_f), \\
\label{rateeq_d}
\frac{d\mathcal{N}_{\mathrm{ph}}}{dt} &=& \Gamma \sum_f N_f
\end{eqnarray}
for a set of lower $\{\ket{i}\}$ and upper $\{\ket{f}\}$ vibronic states and ignoring the spin-rotational structure. The upper states decay with rate $\Gamma$ (assumed to be independent of the vibrational state), and the branching ratio for spontaneous decay from $\ket{f}$ to $\ket{i}$ is denoted as $b_{fi}$. The excitation rate due to the \textit{p}-th laser $R_{fi,p}$ is expressed as
\begin{equation}
    R_{fi,p}=\frac{\Gamma}{2}\frac{b_{fi}I_p/I_{\mathrm{sat}}}{1+4\delta_{fi,p}^2/\Gamma^2},
\end{equation}
where $\delta_{fi,p}$ is the frequency detuning of the \textit{p}-th laser for a certain transition, and $I_p/I_{\mathrm{sat}}$ is the saturation parameter of the laser. 
The laser detuning is set to zero to determine the maximum achievable radiation force, which provides a sufficient result for identifying the required repumping lasers.

The probability of a molecule entering a dark state after scattering several photons follows a geometric distribution \cite{dirosa2004a}. The maximum number of scattered photons is given by $n_{\mathrm{max}}=R_{sc} \, t$, where $R_{sc}  =\Gamma\sum_fN_f$ is the photon scattering rate, and $t$ is the time of interaction between the laser and the molecule. When $n_{\mathrm{max}} \to \infty$, the mean number of photons that the molecule can scatter before leaving the optical cycle tends to $\langle n \rangle \to 1/(1-r)$, where $r$ is the probability of remaining in the optical cycle \cite{hofsass2021a}. For vibrational optical cycling, $r = b_{00}$.

The number of scattered photons required to slow the molecules from 530~m/s to 10~m/s (at which point \Hetwo is at the MOT capture velocity \cite{zenga}) depends on the recoil velocity of the chosen laser cooling transition. For the $a-b$, $a-c$, and $a-e$ cooling transitions, this is approximately $2\times10^4$, $1\times10^4$, and $5\times10^3$ photons, respectively. 
Depending on the cooling transition and the vibrational branching ratios, achieving the required number of scattered photons will necessitate a certain number of repump lasers, as with a single laser, only a mean of 200, 11 and 32 photons will be scattered, respectively. 

The rate-equation simulation is shown in \Fref{fig:OpticalCycling}. In the numerical calculation, the saturation parameters $I_p/I_{\mathrm{sat}}$ for the cooling lasers $C_{00}^{a-c}$ and $C_{00}^{a-e}$ are set to 200 and 100, respectively. 
Increasing the saturation parameter of the cooling lasers beyond these values does not benefit the cooling process, as the transitions are already saturated. 
The saturation parameters for the repump lasers are chosen so that they do not limit the achievable scattering force.  Due to the branching ratios, $I_p/I_{\mathrm{sat}}=1$ is sufficient for the lasers $R_{10,20,30}^{a-e}$ and $R_{21,32}^{a-c}$, whereas $R_{10}^{a-c}$ requires a value of 100. A high saturation parameter for the $a-b$ transition can be easily achieved because of its long lifetime, and therefore, all of the laser saturation parameters are set to 100. With existing laser sources for the relevant wavelengths, we consider the assumed saturation parameters to be viable.

The first two rows show simulation results for a simple approach, where the same electronic transition is used for cooling and repumping. The vibrational repumping for $a-b$ and $a-c$ is achieved by vibrational transitions with $\Delta v=1$, in order to efficiently drive the transitions. For the $a-e$ transition, higher vibrational levels of the excited state are subject to predissociation (see discussion in \Sref{subsec:Predissociation}) and should be avoided during the optical cycling.   

\Fref{fig:OpticalCycling} (a)-(c) shows the number of scattered photons $\mathcal{N}_\mathrm{ph}$ and scattering enough photons to reach 10~m/s (indicated by the horizontal gray dashed line) takes between one and tens of milliseconds, with the latter corresponding to cooling using the $a-b$ transition. While this transition offers highly diagonal branching ratios, requiring only a single repump laser, the long lifetime of the $b$ state makes slowing very inefficient. In contrast, using the $a-c$ and $a-e$ transitions enables more rapid cooling due to higher decay rates and recoil velocities, as shown in \Fref{fig:OpticalCycling} (b)-(c). For $a-c$, additional repumpers are required because of larger off-diagonal branching ratios compared to a-e (see \Fref{fig:BranchingRatios}). 
With a maximum of two repump lasers for $a-c$, and one for $a-e$, both schemes are experimentally viable.

Based on the steady-state solution of \Eref{rateeq_a} and \Eref{rateeq_b}, the maximum scattering rate decreases as the number of lower states coupled to the same upper state increases, scaling with the factor $N_f/(N_i+N_f)$, which can be clearly observed by comparing \Fref{fig:OpticalCycling} (f). This reduction occurs because the population is distributed equally among all coupled states when the scattering rate is saturated. To mitigate this issue, one approach is to repump the population to higher vibrational levels of the upper state, as illustrated in \Fref{fig:OpticalCycling} (d)-(e). Since the population in $a(v\ge2)$ arises solely from spontaneous emission,  most population remains in the main cooling cycle (the states coupled by $C_{00}+R_{10}$), which results in the scattering force being approximately $\hbar k\Gamma/3$.

To prevent the repump lasers from reducing the optical force, a combination of two electronic transitions is investigated, employing one for the cooling and the other for repumping, with the results shown in \Fref{fig:OpticalCycling} (g)-(i). Introducing the second upper electronic state in the optical cycling increases the factor $N_f/(N_i+N_f)$, by reducing $N_i$, resulting in a larger scattering force. When employing the $b$ state as the second electronic state, however, its long lifetime becomes the primary limitation. This is because population accumulation in the $b$ states impedes the quasi-optical cycling efficiency, leading to a reduction of the scattering force to $\approx \hbar k \Gamma/10$.  
The $C:a-c/e,R:a-e/c$ scheme offers the distinct advantage, that both states have a comparable lifetime. Minimal population remains in the repump cycle only due to spontaneous emission, enabling the scattering force to approach the theoretical maximum of $\hbar k \Gamma/2$.

According to this analysis, the optimal scheme for the $a-b$ transition is $C_{00} + R_{10}$, which yields a maximum scattering rate $R_{sc}^{a-b} = \Gamma_{a-b}/3 \approx 5.2 \times 10^5 \, $s$^{-1}$. When repumping through a second electronic state, the scattering rates for $a-c$ and $a-e$ approach the maximum value $\Gamma/2$, which corresponds to $R_{sc}^{a-c} \approx 1.1\times10^7 \, $s$^{-1}$ and $R_{sc}^{a-e} \approx 9.0 \times 10^6 \, $s$^{-1}$, respectively. Rotational branching may reduce the scattering rate, as detailed in the following section.

\subsection{Spin-rotational optical cycling}

\begin{figure}
    \centering
    \includegraphics[width=\columnwidth]{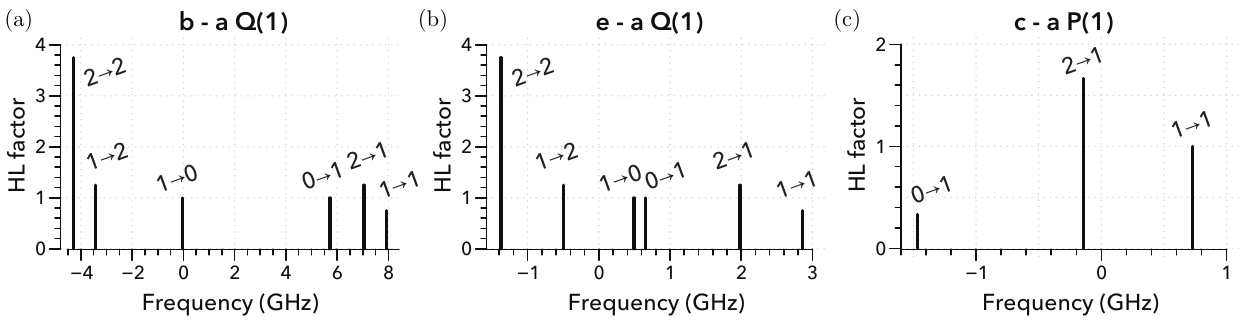}
    \caption{Fine structure components of the $Q(1)$ and $P(1)$ lines of the $^3\Pi$$-$$^3\Sigma$ and $^3\Sigma$$-$$^3\Sigma$ transitions in \Hetwo. The frequency for the spin-free transition was put to zero.}
    \label{fig:StickSpectrum}
\end{figure}

Rotational branching is determined by the selection rules given in \Eref{eq:WET} and \Eref{eq:linestrengthM2S}, and illustrated in \Fref{fig:linestrength}. As illustrated in \Fref{fig:Symmetry}, starting from a $\Sigma$ ground state, closed rotational cycling can be achieved by exciting the $P(1)$ line of a $\Sigma-\Sigma$ transition, or any $Q$ line for a $\Pi-\Sigma$ transition \cite{fitch2021a}. To investigate the effect of the fine structure, \Fref{fig:StickSpectrum} shows the stick spectrum of the $P(1)$ line of the $a-c$ transition and the $Q(1)$ lines for $a-b$ and $a-e$. We note that the fine structure can always be resolved.

Since the rotational state $c(N=0)$ has only one fine-structure component, $J=1$, the spacing between the spectral lines in \Fref{fig:StickSpectrum} (c) reflects the fine-structure splitting of $a(N=1)$. The determination of the Hönl-London (HL) factors using \Fref{fig:linestrength} can be found for this case in \ref{Ap:ExampleHL}. 

For $\Pi$$-$$\Sigma$, six lines appear in the stick spectrum resulting from the three fine-structure components of both $a(N=1)$ and $b/e(N=1)$. The difference of the line splittings in \Fref{fig:StickSpectrum} (a) and (b) can be explained by the scaling of the fine-structure constants with $n$, which leads to a larger splitting between $J$ levels for $b$ than for $e$ (see \Fref{fig:FineStructure}). 

For the $a-c$ $P(1)$ line, there is no possibility that avoids the population of the three lower fine-structure components during optical cycling. This leads to pumping into dark states, if the fine-structure components are individually addressed or mixed using magnetic dipole transitions. Including the Zeeman sublevels, nine for the lower state and three for the upper state, the optical force is reduced to $\hbar k\Gamma/4$ with $R_{sc}^{a-c} \approx 5.6\times10^6 \, $s$^{-1}$. This assumes an optimized angle between magnetic field and laser polarization to destabilize the dark states \cite{berkeland2002a}. 

For the $a-e$ $Q(1)$ line, driving the $J=1\to J'=0$ fine structure component (sometimes in the literature this would be labeled $^QP(1)$ \cite{whiting1972a}) nominally closes the optical cycle. Including the Zeeman sublevels, this would lead to a decrease of the optical force to $\hbar k\Gamma/4$ with $R_{sc}^{a-e} \approx 4.5 \times 10^6 \, $s$^{-1}$.

\begin{figure}
    \centering
    \includegraphics[width=\columnwidth]{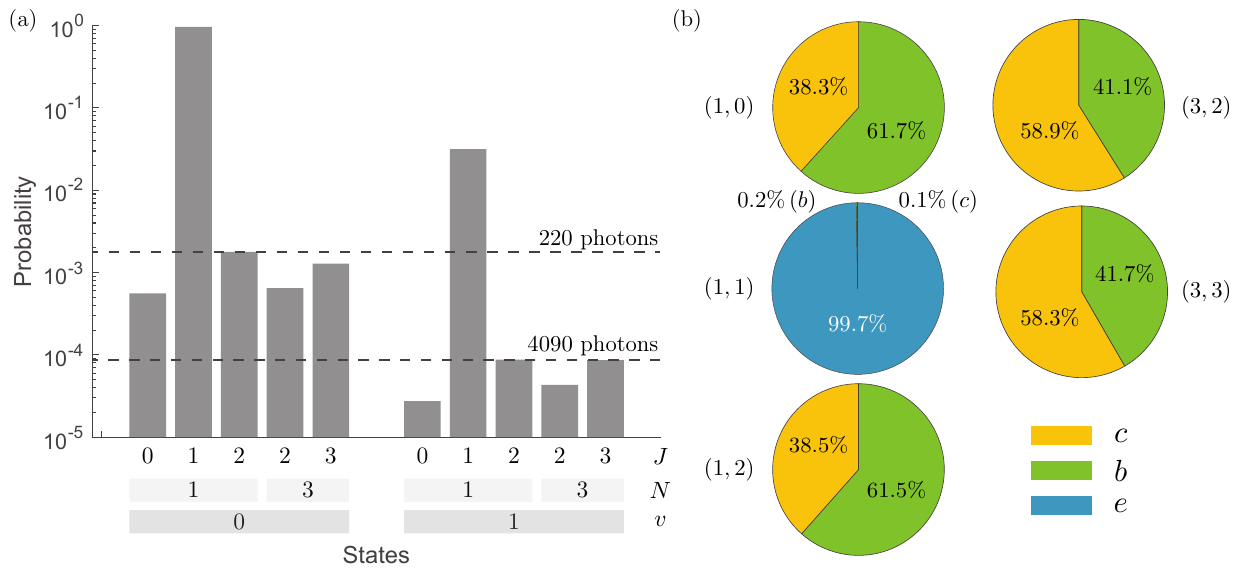}
    \caption{ (a) Population distribution of the $a$ spin-rovibrational levels resulting from the decay of the $e (N=1, J=0, v=0)$ state. (b) Contributions of the direct ($e$) and cascade ($b,c$) decays to the population of the different $a(N,J)$ states. }
    \label{fig:DecayScheme}
\end{figure}

As mentioned previously, the $e (N'=1, J'=0, v'=0)$ level can radiatively decay to the $a$ or $d$ state with the electronic branching ratios given in \Tref{tab:Lifetimes}. Due to the large energy difference of the emitted photon, the direct decay to $a(N=1, J=1)$ is clearly dominant. Only 0.7\% of the molecules will fluoresce to the $d(N=1,J=1)$ state, which further decays to $a$ through $b$ or $c$, causing the population of other spinrotational levels of the $a$ state.
Therefore, after sufficient time, all molecules eventually accumulate in the $a$ state, with the spin-rovibrational population distribution shown in \Fref{fig:DecayScheme} (a).

In the fluorescence cascade, three photons are emitted, setting the maximum accessible total angular momentum to $J = 3$. This explains the absence of population in the $a(N = 3, J = 4)$ state. The same structure can be observed for the $a(v=1)$ levels, scaled by the vibrational branching ratios.

\Fref{fig:DecayScheme} (b) illustrates the relative contributions of the different decay pathways to the population of the fine-structure levels $a(N,J)$. To compute the population distribution in $a$ resulting from the decay cascade of $e$, we sequentially propagate the population through each intermediate state ($d$, $b$/$c$), treating each as a new source of spontaneous emission. This process is governed by the respective vibrational and spin-rotational branching ratios, which determine the relative contributions shown in the pie charts. 

As mentioned in \Sref{sec:Vibronic}, to slow \Hetwo ~molecules using $a-e$, they need to scatter around 4\,850 photons. Considering only one vibrational repumper, $\langle n \rangle = 220$ photons will be scattered before the rotational dark states get populated. By adding rotational repumpers for $v=0$, the mean number of scattered photons increases to 4\,090, making the use of the $a-e$ transition feasible for cooling.

\section{Loss channels} 
\Fref{fig:PES} (b) shows that all excited states - singlet and triplet - lie in the dissociation continuum of the repulsive $X \, ^1\Sigma_g^+$ state. Two factors will affect the dissociation rate of the triplet states in this case: i.) the magnitude of singlet-triplet mixing caused by rotational and relativistic terms in the molecular Hamiltonian and ii.) the overlap of the radial nuclear wave functions of the bound states and the $X$ continuum functions. 

The ultimate fate of \Het is therefore dissociation into the atomic fragments. For the triplet states located below the lowest triplet dissociation threshold, $2~^3S + 1~^1S$, radiative decay to the $a$ state is the dominant decay channel, which dissociates through a spin- and dipole-allowed bound-free transition via the A~$^1\Sigma_u^+$ state.

Above the $2~^3S + 1~^1S$ thresholds, \cc levels with $v>3$ \cite{lorents1989a} are quasi-bound (shape resonances) and predissociate by rotation and for the \bb state, higher vibrational levels can electronically predissociate because of a heterogeneous interaction (Feshbach resonances). These processes do not affect the low-vibrational levels of \bb and \cc states involved in the optical cycling, but the situation for the \ee state is different and will be discussed below. However, \cc states with $v>3$ can be populated via the fluorescence cascade $e\to d\to c$ and we estimate that, after scattering $10^5$ photons on the $a$–$e$ main cooling transition, only 0.1\% of the molecules will be lost through this channel.

\subsection{Spin-forbidden transitions}
\begin{figure}
    \centering
    \includegraphics[width=\columnwidth]{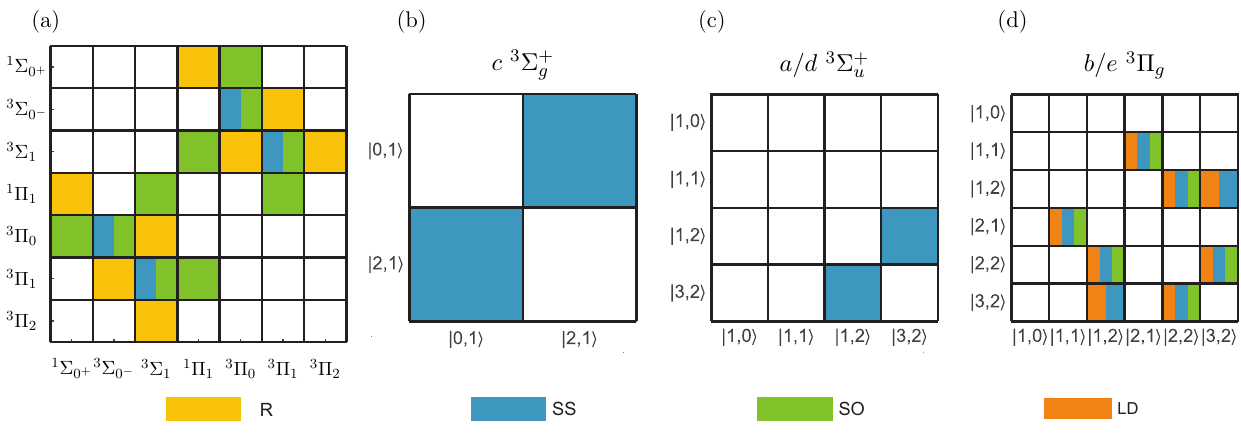}
    \caption{Off-diagonal matrix elements between (a) different electronic states $^{2S+1} \Lambda_{\Omega}$, and different spin-rotational levels $\ket{N,J}$ of (b) $^3\Sigma_g^+$, (c) $^3\Sigma_u^+$, and (d) $^3\Pi_g$ electronic states in \Hetwo. See text for details.}
    \label{fig:Interactions}
\end{figure}

Singlet-triplet mixing in the excited states employed in the optical cycling scheme can lead to spin-forbidden transitions, causing losses to the singlet manifold. The effective Hamiltonian employed in the previous sections accounts for the interactions with other electronic states in terms of the level energies, but does not provide mixing coefficients. 
For a qualitative discussion, \Fref{fig:Interactions} (a) depicts the off-diagonal matrix elements of the molecular Hamiltonian for $\Lambda=0,1$ and $S=0,1$ in a Hund's case (a) basis $^{2S+1} \Lambda_{\Omega}$, where $\Omega$ is the projection of $J$ onto the internuclear axis. States with opposite parity and inversion symmetry cannot mix. 

Rotational perturbations \cite{lefebvre-brion2004b} (indicated by R and in yellow in \Fref{fig:Interactions}) obey the selection rule $\Delta S=0$, and the same holds true for the spin-spin interaction (SS, blue). Moreover, it can be noticed, that none of the $\Sigma$ interact with each other and interactions within the $\Pi$ block are only possible for $^3\Pi_1-^1\Pi_1$, caused by $\hat{H}_\mathrm{SO}$.
In Hund's case (b), $^3\Sigma$ states with $J=0$ will have pure $^3\Sigma_{0^-}$ character at zeroth order, and while they can be perturbed by $^3\Pi_0$ and $^3\Pi_1$ state, no mixing with $S=0$ states is possible. For other $J$ levels, $^3\Sigma_{1}-^1\Pi_1$ mixing caused by $\hat{H}_\mathrm{SO}$ occurs. For the $^3\Pi$ states in Hund's case (b), levels with $J=0$ will have nominally pure $^3\Pi_0$ character and can therefore be perturbed by $^1\Sigma_{0^+}$. For higher $J$, the $^3\Pi_1-^1\Pi_1$ interaction can admix some singlet character.

By comparing the lifetimes of the $A$ ($\tau\sim2$~ns \cite{yarkony1989a}) and $a$ ($\tau\sim18$~s \cite{chabalowski1989a}) states, it can be seen that it is the vanishing singlet-triplet mixing that is responsible for protecting the ungerade excited triplet states from fluorescing into the vibrational continuum of the $X$ state. For the excited gerade state, Yarkony calculated the spin-forbidden transition moments for $A-b$ and $A-c$ and found values on the order of $\sim 10^{-5}$~a.u., resulting in spin-flip branching ratios $<10^{-10}$.
For the $3p\pi$ configuration, the singlet-triplet splitting is $\sim 312.9 \,$cm$^{-1}$ \cite{herzberg1950b}. However, as discussed above, the $e(J=0)$ state mixes only with $^1\Sigma$ states, of which the nearest—$3p\sigma$ and $4p\sigma$—lie several thousand cm$^{-1}$ below/above in energy. We therefore conclude that optical pumping into singlet states is negligible on the timescale relevant to the envisioned experiment.

\subsection{Rotational mixing}
Rotational levels of an electronic state can be coupled by different terms in the effective Hamiltonian (see \Fref{fig:Symmetry}). The off-diagonal matrix elements of $\hat{H}_{\mathrm{eff}}$ in Hund's case (b) are depicted in \Fref{fig:Interactions} (b)-(d). For $\Sigma$ states, the spin-spin interaction is the only one that leads to mixing between rotational states ($\Delta N = \pm 2, \Delta J=0$). As a result, $a(N=1)-c(N'=0)$ transition is not completely closed due to the $N=3$ character of the lower $a(N=1, J=2)$ level and the $N=2$ character of the upper $c(N'=0,J'=1)$ level. However, the interactions between these states are sufficiently weak that an average of $10^7$ photons must be scattered before populating $a(N=3)$. For $\Pi$ states, both spin–orbit coupling and $\Lambda$-doubling give rise to interactions between levels with different rotational quantum number $N$.
In this case, all values of $N$ are permitted by the Pauli principle due to the presence of the $\Pi^\pm$ doublet. These two components are further coupled through the spin–spin ($H_{SS}$), spin–orbit ($H_{SO}$), and $\Lambda$-doubling ($H_{LD}$) interactions. For the $a(J=1)-e(J'=0)$ transition the involved levels remain unperturbed.

\subsection{Predissociation of the excited states} \label{subsec:Predissociation}
The bound vibrational states of $a$, $b$, and $c$ lie below the lowest dissociation limits associated with a triplet state and predissociation plays no role. The lowest vibrational levels of the $d$ and $e$ states, however, lie in the $a$ and $c$ state dissociation continuum. With $d$ and $a$ possessing the same symmetry, predissociation can occur because of vibronic interactions. For the $e(v'=0)$ states, predissociation is possible only into the $c~^3\Sigma_g$ continuum, while for higher $v$ the $b~^3\Pi$ continuum also becomes accesible. We use Fermi's Golden Rule to obtain an order-of-magnitude estimate of the predissociation rates of $d(v'=0)$ and $e(v'=0)$. 

According to \Fref{fig:Interactions}, $e \, ^3\Pi_g^+$ can interact with $c \, ^3\Sigma_g^+$ via different couplings, enabling predissociation. Since $A_{SO}/B \ll 1$, the dominant interaction comes from the rotational term and the associated operator corresponds to the gyroscopic perturbation, $-\frac{1}{2\mu R^2}(J_+L_- + J_-L_+)$ \cite{lefebvre-brion2004b}. Using the pure precession approximation to compute the matrix element between bound and continuum states, an estimate of $10^4 \, \mathrm{s^{-1}}$ is obtained for the predissociation rate of $e(v'=0)$. This is three orders of magnitude smaller than the radiative rate (see \Tref{tab:Lifetimes}), implying that a molecule can scatter 1\,000 photons before it dissociates. To minimize losses during optical cycling, it is therefore desirable to avoid populating the $\Pi^+$ component of the $e$ state. The $e(J'=0)$ level employed in the $a-e$ cooling cycle remains unperturbed and the pure $\Pi^-$ character prevents dissociation into the $c$ state continuum.

Molecules in the $d \, ^3\Sigma_u^+$ state can be lost by nonadiabatic predissociation, related to off-diagonal elements of the $d/dR$ operator. Because neither theoretical nor experimental data for this process is available, nonadiabatic coupling elements for the corresponding $2s-3s$ states in H$_2$ at $2a_0$ \cite{quadrelli1990b}. Such an estimate using Fermi's Golden Rule results in an upper limit for the predissociation rate of $6 \times 10^5 \, \mathrm{s^{-1}}$. We expect the real value to be significantly smaller, because configuration interactions related to the ion-pair state are absent in \Het, while they are prominent in the double-well states of H$_2$. By comparing this value to the radiative rate of the $d$ state, we find that predissociation of the $d$ state is probably negligible under the current conditions.

\subsection{Two-photon ionization}
Two-photon ionization is not energetically possible when driving the $a-b$ or $a-c$ cooling transitions, as can be seen from \Fref{fig:PES} (b). However, for the $e \, (v'=0)$ state, located 21\,507.25$\,$cm$^{-1}$ \cite{brown1971b} above the $a \, (v=0)$ state, absorption of a second photon allows to reach the ionization continuum (ionization energy $34\,301.207\,$cm$^{-1}$ \cite{semeria2020a}), creating \Hep in its electronic ground state $X^+ \, (v^+) $ and a free electron with energy $ h \nu - U_I(v^+)$.
 Here, $U_I(v^+)$ is the ionization energy of $e(v' = 0)$ with respect to the limit $X^+(v^+)$ and $\nu$ is the frequency of the photon.

The two-photon ionization rate is given by $R_I = \sigma_I I/(h \nu)$, with $\sigma_I$ being the photoionization cross section and the figure-of-merit is the ratio of ionization versus radiative decay, which can be written as $b_I = R_I / \Gamma$ for $\Gamma\gg R_I$.
The ionization cross section (in cm$^2$) for hydrogen-like atomic systems is given by \cite{chupka1987a}
\begin{equation}
    \sigma = \frac{8 \times 10^{-18}}{Z \, (U_I/\mathcal{R})^{1/2} \, (h\nu/U_I)^3},
\end{equation}
where $Z$ is the ion charge and $\mathcal{R}$ is the Rydberg constant. To adapt this equation to the molecular case, we multiply with the $e-X^+$ Franck-Condon factor to account for the vibrational excitation during ionization.

From a vibrational perspective, the transition between $v' = 0$ and $v^+ = 0$ is the most likely. However, the photon energy is significantly higher than the ionization potential, which reduces the probability of electron escape. The opposite holds true for the highest reachable ionization continuum with $v^+ = 5$: the photon energy is closer to the ionization energy, increasing the likelihood of ionization, but the vibrational transition is highly suppressed.
Using this approach, we obtain an order-of-magnitude estimate of the two-photon ionization cross section for $a-e$ and the corresponding loss rate: $\sigma_I \approx 5 \times 10^{-18}\,$cm$^2$ and $R_I = 4.3\,$s$^{-1}$ (assuming $I/I_{\mathrm{sat}} = 100$). The resulting photoionization branching ratio, $b_I = 2.4 \times 10^{-7}$, is sufficiently low that it does not impose any constraint on the required number of photon scattering events.

\subsection{Penning ionization}
Collisions between metastable \Het molecules can lead to Penning ionization \cite{eltsov1995a} according to
\begin{eqnarray}
\mathrm{He}^*_2\left(a~^3\Sigma_u^{+}\right) + \mathrm{He}^*_2\left(a~^3\Sigma_u^{+}\right) & \to 3 \mathrm{He}\left(1~^1S\right) + \mathrm{He}^{+}\left(1~^2S\right) + e^{-} \\
& \to 2 \mathrm{He}\left(1~^1S\right) + \mathrm{He}_2^{+}\left(X^+~^2\Sigma_u^{+}\right) + e^{-},
\end{eqnarray}
a barrier-less process that proceeds with near-unit probability at close range.
As in the case of mutual recombination of metastable atomic He($2~^3S$), spin conservation governs the reaction dynamics. When the molecular spins are aligned, the initial complex forms a quintet state, whereas the reaction products are restricted to singlet or triplet configurations.
This allows for the suppression of Penning ionization if the molecules are prepared in the spin-stretched state.

\section{Conclusion and Outlook}

We have shown that the metastable helium dimer presents a promising candidate for direct laser cooling, with three accessible cycling transitions spanning the infrared, near-infrared, and ultraviolet spectral regions.
Rate-equation simulations indicate that sufficient cycling efficiency can be achieved to enable the scattering of approximately $>10^4$ photons, enabling laser slowing and magneto-optical trapping.
Moreover, the availability of multiple electronic transitions with near-diagonal Franck–Condon factors allows for vibrational repumping without compromising the maximum achievable optical force.

We discussed the situation for optical cycling in homonuclear molecules, where the Pauli-principle restricts the existence of some rotational levels. The effects of off-diagonal terms in the effective rotational Hamiltonian was analyzed, also taking into account the cascading fluorescence from the $e$ state. As for the estimated predissociation lifetimes of the $e$ and $d$ states, experimental measurements of the lifetimes and branching ratios are required to verify the proposed schemes.

Ultracold samples of \Het will facilitate precision measurements of the energy-level structure of \Hetwo and \Hep through increased interaction times and a suppression of Doppler-related effects. As a homonuclear diatomic cation, \Hep has no electric-dipole allowed rovibrational transitions, but its energy-level structure is accessible via MQDT-assisted Rydberg extrapolation: (i) molecular Rydberg series of the neutral dimer, converging to different rovibrational states of the ion, are excited from a common ground state; (ii) the ionization limits are obtained by extrapolating the experimentally obtained term values for a collection of Rydberg states to principal quantum number $n \to \infty$ using MQDT, which takes nonadiabatic interactions between Rydberg series into account; and (iii) rovibrational intervals in the ion are obtained by subtracting the corresponding ionization energies, obtained from a common initial state in the neutral molecule. 
This procedure will allow to bechnmark the results from QED calculations for this fundamental molecular three-electron system. 

\begin{figure}
    \centering
    \includegraphics[width=0.5\linewidth]{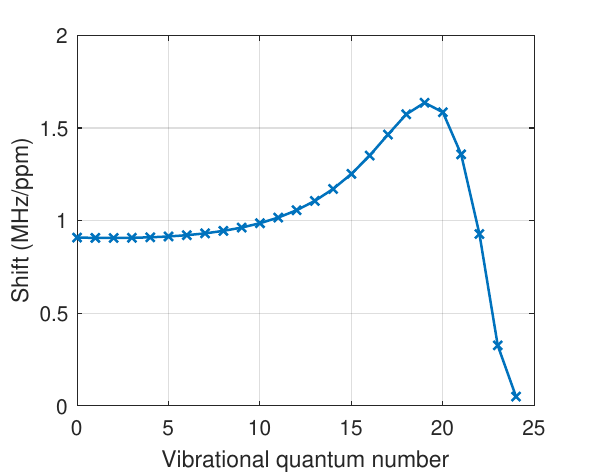}
    \caption{Shift of vibrational energy levels in MHz for a 1$\,$ppm change of the static polarizability.}
    \label{fig:PolarizabilitySensitivity}
\end{figure}

Additionally, long interaction times in combination with optical pumping allow the preparation of excited vibrational levels, giving access to the Rydberg states converging to the highest vibrational levels of \Hep. This allows to use \Hep as a molecular quantum sensor to determine the static dipole polarizability of atomic helium in the $^1S_0$ state, for which very accurate theoretical predictions exists, but which is difficult to measure spectroscopically. 
Induction and dispersion forces control the long-range region of the He$^+$- He interaction, and give rise to an attractive interaction of the form \cite{kaplan2006a}
\begin{equation}
    V(R) = - \frac{\alpha_{\mathrm{He}}}{2R^4} - \alpha_{\mathrm{FS}}^2\frac{W_4}{R^4} - \mathcal{O}(R^{-6}) \qquad \mathrm{for} ~ R \to \infty,
    \label{eq:alphaPotential}
\end{equation}
where the second term is a relativistic correction \cite{meath1966a} and $\alpha_{\mathrm{FS}}$ is the fine-structure constant. Using an analytical potential energy curve for \Hep \cite{xie2005a}, the effect of varying $\alpha_{\mathrm{He}}$ on the level energies can be investigated by numerically solving the radial Schrödinger equation. The result is shown in \Fref{fig:PolarizabilitySensitivity}, which gives the shift of the rotation-less ($N = 0$) levels with vibrational quantum number $v^+$ for a 1$\,$ppm change of the electric dipole polarizability of atomic He, $\alpha_{\mathrm{He}} = 1.38376077(14) \, a_0^3$ \cite{puchalski2020b}. The maximum shift of about 1.6$\,$MHz occurs for the level $v^+ = 19$, while levels
with $v^+ < 10$ shift by about 0.9$\,$MHz.

The measurement of intervals between excited vibrational states with an accuracy of better than 100$\,$kHz holds the promise of testing the calculated polarizability at the level of its theoretical uncertainty. 

The dipole polarizability is also relevant for new quantum pressure standards. These new standards rely on measuring the refractive index or the electric permittivity of an atomic gas, along with the polarizability of the atom obtained from ab initio calculations. The key is the (ideal) gas law, which relates the pressure $p$ to the particle density $\rho = N/V = \rho_m N_A$,
\begin{equation}
    p = \frac{N}{V} \, k_B T = \rho k_B T = \rho_m R T,
    \label{eq:IdealGas}
\end{equation}
with $k_B$ the Boltzmann constant, $R$ the ideal gas constant, $N_A$ the Avogadro's number, and $T$ the temperature. The working equation for dielectric-constant gas thermometry (DCGT) is the Clausius-Mossotti equation (for simplicity ideal gas behavior is assumed here, see \cite{gaiser2015a} for a general treatment),
\begin{equation}
    \frac{\epsilon_r - 1}{\epsilon_r+2} = \frac{4\pi}{3} \alpha \rho,
    \label{eq:DCGT}
\end{equation}
indicating the particular importance of the polarizability $\alpha$. Together, \Eref{eq:IdealGas} and \Eref{eq:DCGT} allow the realization of a pressure standard by measuring the dielectric constant of the gas ($\epsilon_r$) via the capacitance of a gas-filled capacitor, which was recently demonstrated \cite{gaiser2020a}. Currently, the best experimental values for rare gas polarizabilities were obtained using DCGT and were based on traditional pressure standards, giving a value of $\alpha_{\mathrm{He}} = 1.38376162(27) \, a_0^3$ \cite{gaiser2018a}. Instead of relying on DCGT for both testing ab initio polarizabilities and establishing a new standard, these two processes can be disentangled using an independent approach to experimentally determine $\alpha$, based on an ultracold gas of \Het and Rydberg spectroscopy thereof.

\section{Acknowledgment}

We thank Frédéric Merkt, Paul Jansen, and Maxime Holdener (ETH Zurich) for useful discussions. 

The work was supported by the European Union (ERC, HeliUM, 101116599) and the Dutch National Growth Fund (NGF), as part of the Quantum Delta NL programme (NGF.1623.23.016).
Views and opinions expressed are however those of the author(s) only and do not necessarily reflect those of the European Union or the European Research Council Executive Agency. Neither the European Union nor the granting authority can be held responsible for them.

\clearpage

\appendix

\section{Potential energy curves}

For the calculation of FC factors and radiative lifetimes, the BO potential energy curves (PEC) for the a, b and c state were taken from Ref.~\cite{yarkony1989a} and rovibrational energy levels were calculated using the DVR method given by Colbert and Miller \cite{colbert1992a}. While these curves result in rovibrational intervals that agree with the available experimental data at the level of a few \wn, the electronic energies show deviations of the order of hundred \wn, indicating a correct shape of the PEC and an approximately $R$-independent offset. To correct for that, we shifted the PEC of the a state relative to the PEC of the $X^+$ state of \Hep, to bring the calculated ionization energy in agreement with the experimental value of 34\,301.207\,00(4)~\wn \cite{semeria2020a}. Subsequently, the PEC of the b and c state were shifted so that the 0-0 band origin agreed with the experimental values given by Focsa \cite{focsa1998a}.  

In lack of available ab initio data, the PEC of the $d(3s\sigma)$ Rydberg state was taken to be the one of $X^+$ shifted to reproduce the 0-0 band origin of $a-d$ in ~\cite{ginter1965e}. For the $e(3p\pi)$ Rydberg state, taking into account the $R$-dependent $\eta_{p p}^{(\Pi)}$ quantum-defect functions obtained by Sprecher and coworkers \cite{sprecher2014b}, the PEC was obtained using \Eref{eq:PEC} and by solving the QDT quantization condition 
\begin{equation}
\frac{\tan \pi \sqrt{\frac{-1}{\varepsilon(R)}}}{A_{\ell=1}(\varepsilon)}+\tan \pi \eta_{p p}^{(\Pi)}(R, \varepsilon)=0
\end{equation}
for each $R$, with $A_{\ell=1}=1-\varepsilon(R)^2$ . After shifting the PEC to reproduce the 0-0 band origin of $a-e$ in \cite{brown1971b}, transition frequencies agree at a level of 1~cm$^{-1}$ with the experimental observations reported by \cite{brown1971b}.

\section{Electronic transition moments}
\label{sec:ElectronicTransitionMoments}
Transition moments for the $a-b$ and $a-c$ transitions were calculated by Yarkony \cite{yarkony1989a} and with the classical turning points for the involved states falling within [1.8-2.5]$a_0$, the transition moments vary by about 6\% and 4\%, respectively.
Given the small $R$-depedence, the missing transition moments were obtained at $R_{eq}=2~a_0$ using the Coulomb approximation \cite{bates1969b}. From the experimental electronic Term values $T_e$, the effective principal quantum numbers were determined to be $n^*_a\approx 1.789, n^*_b\approx 1.928, n^*_c\approx 2.170, n^*_d\approx 2.813$ and $n^*_e\approx 2.933$. 

The molecular electronic transition moments can be obtained using $\mu_q=-e r C^{(1)}_q$ with $C^{(1)}$ being the normalized spherical harmonic, and by applying the Wigner-Eckart theorem in the molecular frame
with $\braket{L' || C^{(1)} || L} = (-1)^{L_>+L'}\sqrt{L_>}$ and $L_>=max(L',L)$. For $\mathrm{s}\sigma\--\mathrm{p}\sigma$ and $\mathrm{s}\sigma\--\mathrm{p}\pi$ the angular factor is $\sqrt{1/3}$. The radial matrix elements $\braket{\eta' || r || \eta}$ were calculated by integrating the radial Schrödinger equation at the energies given by the effective quantum number. The wave functions were obtained using the renormalized Numerov method, integrating inwards and stopping the integration at $r_\mathrm{core}=0.5~a_0$.

The transition moments in atomic units are: $\mu_{ab}=-2.65, \mu_{ac}=3.02, \mu_{ae}=0.943, \mu_{db}=0.837, \mu_{dc}=-2.19$ and $\mu_{de}=-6.84$.

\section{Transition strengths}
\label{Ap:ExampleHL}

As an illustration on how to use \Fref{fig:linestrength} to determine absolute and relative line strengths, we consider the case of emission and absorption for the $a-c$ transition. 
\vspace{5pt}

\textit{Emission:} $\Sigma \to \Sigma$ \\
The $c(N=0,J=1)$ state can only decay to the $a(N=1)$ levels (P(1) transition). To find the relative intensities of the three fine-structure components, we use \Fref{fig:linestrength} (c). According to the first row, which corresponds to a transition from 0 to 1, we get that $\frac{33.3 \, \cdot \, 100}{300} = 11.1\%$ of the intensity goes to $J=0$, $\frac{100 \, \cdot \, 100}{300} = 33.3\%$ to $J=1$, and $\frac{167 \, \cdot \, 100}{300} = 55.6\%$ to $J=2$. 

To determine the absolute values, we also need \Fref{fig:linestrength} (b). Our case of interest corresponds to a $\Sigma \to \Sigma$ transition from $N'$ to $N'+1$, with $N'=0$. Taking the values from the same row as before, we now find that the line strength factors are $\frac{33.3}{100} \, \cdot \, \frac{N'+1}{2N'+1} = 0.333 \, \cdot \,  1 = 0.333$ for $J=0$, $\frac{100}{100}  \, \cdot \, 1= 1$ for $J=1$, and $\frac{167}{100} \, \cdot \, 1 = 1.67$ for $J=2$.

\textit{Absorption:} $\Sigma \leftarrow \Sigma$ \\
According to \Fref{fig:linestrength} (a), there are two possible transitions for an excitation from $a(N=1)$: P(1) to $c(N'=0)$ with a probability of $P_{\mathrm{P}}= \frac{N}{2N+1} = 1/3$; and R(1) to $c(N'=2)$ with $P_{\mathrm{R}} = \frac{N+1}{2N+1} = 2/3$.

The line strengths for the fine structure components are obtained by divided by 100 the values in \Fref{fig:linestrength} (c) and multiplying them by $P_{\mathrm{P}}$ or $P_{\mathrm{R}}$ in each case. The following case corresponds to the R(1) transition, where the values of the 1-2 block have been scaled by $\frac{1}{100} \cdot \frac{2}{3}$ to get the line strengths.

\begin{eqnarray*}
     1 \leftarrow 0: \quad 0.222 &\qquad \qquad 1 \leftarrow 1: \quad 0.167 \qquad \qquad 2 \leftarrow 1: \quad 0.5 \\
     1 \leftarrow 2: \quad 0.011 & \qquad \qquad 2 \leftarrow 2: \quad 0.167 \qquad \qquad 3 \leftarrow 2: \quad 0.933
\end{eqnarray*}

\section{Vibronic branching ratios} \label{Ap:VibBr}

\Tref{tab:AppendixBranchingRatios} shows the vibronic branching ratios computed by considering $R$-dependent electric dipole moments.

\begin{table}[b!]\tiny \caption{\label{tab:VibBr}Vibronic branching ratios.} 
\lineup
\begin{tabular}{l | rrr | rrr | rrr | rrr} \toprule
       & \multicolumn{3}{c}{a-c} & \multicolumn{3}{c}{a-b} & \multicolumn{3}{c}{a-e} & \multicolumn{3}{c}{a-X$^+$}   \\
 \diagbox{$v$}{$v'$} &          0 &          1 &          2 &          0 &          1 &          2 &          0 &          1 &          2 &          0 &          1 &          2  \\\midrule
  0 &   9.12[-1] &   2.02[-1] &   3.88[-2] &   9.95[-1] &   3.47[-2] &   3.42[-3] &   9.68[-1] &   4.78[-2] &   4.00[-3] &   9.51[-1] &   6.09[-2] &   5.51[-3]   \\
  1 &   8.62[-2] &   6.24[-1] &   3.44[-1] &   5.00[-3] &   9.55[-1] &   5.96[-2] &   3.17[-2] &   8.90[-1] &   8.48[-2] &   4.88[-2] &   8.44[-1] &   1.07[-1]   \\
  2 &   1.77[-3] &   1.67[-1] &   3.67[-1] &   1.63[-6] &   9.90[-3] &   9.22[-1] &   7.21[-6] &   6.20[-2] &   8.21[-1] &   1.12[-4] &   9.52[-2] &   7.48[-1]   \\
  3 &   1.35[-5] &   6.60[-3] &   2.34[-1] &            &   8.92[-6] &   1.47[-2] &   8.87[-7] &   6.66[-6] &   9.05[-2] &   1.88[-6] &   2.64[-4] &   1.39[-1]   \\
  4 &   4.94[-9] &   7.71[-5] &   1.65[-2] &            &    &   2.84[-5] &   4.70[-9] &   4.07[-6] &   4.87[-8] &   1.58[-9] &   9.54[-6] &   3.76[-4]   \\\bottomrule
\end{tabular}
\label{tab:AppendixBranchingRatios}
\end{table}

The value of the reduced mass in atomic units is $\mu_{\mathrm{He}_2} = m_{\alpha}/2 = 3647.14977071$ using CODATA 2018 \cite{tiesinga2021a}.

\clearpage
\section*{References}

\bibliographystyle{unsrt}

\end{document}